\documentstyle[sprocl]{article}

\newcommand{\lla}{\langle \langle}
\newcommand{\rra}{\rangle \rangle}
\renewcommand{\d}{\partial}
\newcommand{\beq}{\begin{eqnarray}}
\newcommand{\eeq}{\end{eqnarray}}
\newcommand{\sbeq}{\begin{subeqnarray}}
\newcommand{\seeq}{\end{subeqnarray}}

\newcommand{\do}{^{\circ }}

\newcommand{\btem}{\bibitem}

\newcommand{\e}{{\rm e}}
\newcommand{\TH}{T. Hatusda\ }
\newcommand{\TK}{T. Kunihiro\ }
\newcommand{\PR}{Phys. Rev. }
\newcommand{\NPA}{Nucl. Phys. {\bf A}}

\def\be{\begin{equation}}
\def\ee{\end{equation}}
\def\bea{\begin{eqnarray}}
\def\eea{\end{eqnarray}}

\begin{document}

\title{Chiral Transition and Baryon-number Susceptibility 
\footnote{The title is slightly changed from the original one
``Baryon-number Susceptibility and Related Problems''.}}

\author{Teiji Kunihiro}

\address{Faculty of Science and Technology, 
 Ryukoku University, Seta, Otsu, Japan\footnote{The 
 permanent address after April 1, 2000: 
Yukawa Institute for Theoretical Physics, Kyoto University,
Sakyoku, Kyoto  606-8502, Japan.\ \ E-mail: kunihiro@yukawa.kyoto-u.ac.jp} 
}

\maketitle\abstracts{
We discuss the baryon-number susceptibility $\chi _B$ 
 and related topics which include the density fluctuations
 around the critical point of the chiral transition at
finite temperature $T$ and baryon density $\rho_B$.
 Phenomenological implications of the density fluctuations near the
 first-order chiral transition at finite density are also discussed.}

\section{Introduction}
 
When exploring a phase transition  in any physical 
system, the study of fluctuations of physical quantities, especially ones related
to the order parameter is as important as that of the phase diagram for the
system  in equilibrium. The fluctuations of observables are related with 
dynamical phenomena including the transport properties of the system.
The chirally restored and deconfined phase is expected to be created
 dynamically in the
intermediate stage of the ultra-relativistic heavy ion collisions and
in the early universe. Therefore the study of the fluctuations has
 a great relevance to phenomenology.

In the present report, we discuss the baryon-number susceptibility $\chi _B$ 
\cite{qnum,qnumk} and related topics which include the density fluctuations
 around the critical point of the chiral transition at
finite temperature $T$ and baryon density $\rho_B$.

\setcounter{equation}{0}  
\section{QCD phase diagram in $(T, \rho_B)$-plane and the vector coupling}

The lattice simulations of QCD \cite{lattice,lattice2} suggest that 
the order and even the
  existence  of the phase transition(s) at finite temperature $T$ 
are largely dependent on the 
number of the active flavors when the physical current 
quark masses  are used: 
For $m_u\sim m_d\sim 10 {\rm MeV} <100{\rm MeV}\stackrel {<}{\sim}m_s$,
 the phase transition may be weak 1st order or 2nd order or not exist. 
The lattice QCD is, unfortunately, 
still not matured enough to predict a definite
thing about the phase transition at finite baryon density
 $\rho_B$(or chemical potential $\mu$).

Low-energy effective models\cite{physrep,instanton}
 and the chiral random-matrix theory\cite{random} have  given 
suggestive pictures of the phase diagram 
of QCD in the $T$-$\mu$ (or $T$-$\rho_B$)
 plane. For example, the NJL model\cite{NJL} well describes
 the gross features of the $T$ dependence of the quark condensates of the 
lightest three quarks as given 
by the lattice QCD, and predict that the chiral transition at $\mu\not=0$
is of rather strong first order at low temperatures $T$, say, lower than
 50 MeV, provided that the vector coupling between the quarks as given by 
$g_{_V}/2\cdot (\bar{q}\gamma _{\mu}q)^2$ is absent\cite{asakawa,npb91}.
\footnote{
The first order chiral transition in the density
direction at low temperatures is also obtained in the chiral
random-matrix theory.\cite{random}}
 As a matter of fact, the strength and even the existence of the
 1st order transition are strongly dependent on the strength of the 
 vector coupling $g_{_V}$ \cite{asakawa,npb91}; the vector term prevents a
  high-density state. 

The reason why the vector coupling weakens the phase
 transition and postpones  the 
chiral restoration is understood as follows.
Thermodynamics tells us that when  two phases I and II are coexistent, 
their temperatures $T_{\rm I, II}$, 
pressures $P_{\rm I, II}$ and
 the chemical potentials $\mu_{\rm I, II}$ are the same;
\beq
T_{\rm I}=T_{\rm II}, \quad \  
P_{\rm I}=P_{\rm II}, \quad \  
\mu_{\rm I}=\mu_{\rm II}.
\eeq
If the phase I (II) is the chirally broken (chirally restored) phase,
the last equality further tells us something because $\mu_{\rm I, II}$ 
 at vanishing temperature are given
  by $\mu_{\rm i}=\sqrt{M^2+p_{F_i}^2}$,
 ($i=$ I, II),
where  $p_{F_i}$ is the Fermi momentum of the $i$-th phase,
 and $M$ and $m$ are the constituent quark mass and the current quark mass that 
vanishes in the chiral limit. One readily sees that $p_{F_I}<p_{F_{II}}$, 
accordingly, $\rho_{B_I}<\rho_{B_{II}}$, i.e.,
the chirally restored phase is in  higher density than
 the coexistent broken phase.
The vector coupling 
above gives rise to a repulsion proportional to the density squared i.e.,
$g_{V}\rho_B^2/2$ which is bigger in the restored phase than in the 
broken phase. Thus the vector coupling weakens and/or postpone the phase
transition of the chiral restoration at low temperatures. 

Since the vector coupling contribute to  the energy repulsively,
 it also  suppress the density fluctuations, or the baryon-number 
susceptibility, which
is the main subject of the present report.

\setcounter{equation}{0}  
\section{Baryon-number susceptibility}

The baryon-number susceptibility $\chi_B$  
is the measure of the response of the baryon
 number density $\rho_B=\sum _{\i=1\sim N_f}\rho_i$ 
to infinitesimal changes in the quark chemical potentials $\mu _i$
\cite{qnum,qnumk}:
\beq
\chi _B(T,\mu)=\bigl[\sum_{i=1}^{N_f}\frac {\partial} {\partial \mu_i}
\bigr]( \sum_{i=1}^{N_f}\rho_i)
=\lla {N}_B^2\rra/VT,
\eeq
where $ {N}_B $ is the baryon-number operator 
given by
$ N_B\equiv \sum_{i=1}^{N_f} N_i,$
with
 \beq
\rho _i={\rm Tr} N_i\exp [-\beta( H-\sum _{i=u,d}\mu _i
 N_i)]/V\equiv\lla N_i\rra/V
\eeq
 the $i$-th quark-number density,
 $V$ the volume of the system and $\beta=1/T$. 
 
In the following, we shall confine ourselves to the  
$SU_f(N_f)$-symmetric case;
 $\mu_u=\mu_d=\mu_s=...\equiv \mu.$

  The baryon-number susceptibility 
at $\rho_B\not=0$  is 
 related with the (iso-thermal) compressibility of the system
\cite{qnumk} 
$\kappa _{_{T}}\equiv -N_B^{-1}(\d V/\d\mu )_{T,N_B}
=\frac {\chi_B}{\rho ^2}$, which
 tells  
that if $\chi_B$ is large and so is the density fluctuation, 
the system is easy to  compress.

Another physical meaning  of $\chi_B$ is 
that it is the density-density correlation which is
 nothing but the 
 0-0 component of the vector-vector  correlations or fluctuations\cite{qnumk};
\beq
\chi _B(T,\mu_q)=\beta \int d{\bf x}S_{00}(0,{\bf x}),
\eeq
 where
$S_{\mu \nu}(t,{\bf x})=\lla j_{\mu}(t,{\bf x})j_{\nu}(0,{\bf 0})\rra,$
with
 $j_{\mu}(t,{\bf x})= \bar {q}(t,{\bf x}) \gamma _{\mu}q(t,{\bf x})$
being the  current operator.
Using the fluctuation-dissipation theorem, one has
\beq
\chi _B(T,\mu_q)=-\lim _{k\to 0}L(0,{\bf k}),
\eeq 
where $L(\omega, {\bf k})$ is the longitudinal component 
of the retarded Green's function or the response function in the
 vector channel;\\ 
$R _{\mu \nu}(\omega,k)={\rm F.T.}(-i\theta(t)\lla[j_{\mu}(t,{\bf x}),
\ j_{\nu}(0,{\bf 0})]_{-}\rra)$.

The  lattice simulations of QCD \cite{got}
  show  that $\chi _B$ at $\mu =0$ is suppressed in the  low temperature
phase,    increases with $T$ 
 sharply around the critical point of the chiral transition and takes almost 
the free quark gas value then saturates. This behavior may be understood 
intuitively and roughly as follows\cite{lattice}:
In the confined phase at low $T$, the density fluctuation picks up 
 the  Boltzmann factor $\e^{-M_N/T}$ with $M_N$ being the nucleon mass, 
 which is much smaller than the factor $\e^{-M_q/T}$ with $M_q$
 being the current  quark mass (constituent quark mass), which factor will be
 picked up in the  deconfined and chirally restored (chirally broken) phase.

Our point here is however that the nature of the chiral transition and also
presence (or absence) of the vector coupling affect the
baryon-number susceptibility, especially when $\rho_B\not=0$, for which 
the lattice data is not available so far\cite{qnumk}.

\subsection{Free quark gas}

It is instructive to examine $\chi_B(T, \mu)$ in the  
simple free-quark gas model:
 \beq
\rho_B=
2N_fN_c\int \frac {d{\bf p}}{(2\pi)^3}(n(T,\mu)-\bar{n}(T,\mu))
\eeq
where
$n(T,\mu)=1/[\exp \beta(E_p-\mu)\, +\, 1]$ and 
$\bar {n}(T,\mu)=1/[\exp \beta(E_p+\mu )\, +\, 1]$
 with $E_p=\sqrt{M^2+p^2}$,  and $N_c=3$ is
 the number of the colors.
Then one readily obtains
\beq
\chi_B(T,\mu)= 2N_fN_c\beta \int {{d{\bf p}}\over {(2\pi)^3}}
\Bigl\{ n(1-n)+\bar {n}(1-\bar {n})\Bigr\}
\equiv \chi_B^{(0)}(T,\mu),
\eeq
 which is reduced to 
\beq
\chi_B^{(0)}(T,0)\equiv \chi_B^{(0)}(T)=
4N_fN_c\beta \int {{d{\bf p}} \over {(2\pi)^3}}
{{\exp (E_p/T)}\over {[\exp (E_p/T)+1]^2}},
\eeq
at $\mu =0$.

If $M(T)$ is decreased as in the chiral restoration,
$\chi_B$ increases  and
 reaches $N_fT^2$ at $M(T)=0$: The enhancement is,however, 
 found to be  modest and not so large as obtained in the lattice simulations.

\subsection{Model calculation}

To demonstrate the relevance of the nature of the chiral transition and
 the presence of the vector coupling to $\chi_B$,
 we perform a  calculation with an effective model, which is given by 
 adding the vector-coupling terms \cite{kocic} to 
 the  Nambu$-$Jona-Lasinio model\cite{qnumk}: 
\beq
{\cal L}& =&  \bar{q}(i \gamma \cdot \partial -{\bf m})q + 
  \sum^{N_f^2-1}_{a=0}{g_{_S} \over 2}[(\bar{q}\lambda _a q)^2 + 
  (\bar{q}i\lambda_a \gamma_5q)^2] \nonumber \\
 &\ \ \ \ \ &  - 
 {g_{_V} \over 2}\sum ^{N_f^2-1}_{a=0}
[(\bar{q}\lambda_a\gamma _{\mu } q)^2 +
 (\bar{q}\lambda_a\gamma _{\mu }\gamma_5q)^2]
\eeq
The realistic value of $g_{_V}$
  used in the literature \cite{physrep} is
  roughly in the range   of $g_{_V}\Lambda ^2=5\sim 9$.  

In the self-consistent  mean field approximation, the 
constituent mass (dynamical mass) $M_i(T, \mu_q)$,
 the quark condensates $\lla \bar {q}_iq_i\rra$
 and the quark density $\rho_i$ are all coupled 
 with the vector coupling $g_{_V}$ and determined by the following
equations;
 \beq
M_i&=&m_i-2g_{_S}\lla \bar{q}_iq_i\rra, \\
\lla\bar {q}_i q_i \rra&=&
-2N_c\int \frac {d{\bf p}}{(2\pi)^3}\lbrace 1-n_i(T,\tilde {\mu}_i)
-\bar {n}_i(T,\tilde {\mu}_i)\rbrace, \\
\rho_i&=&
 2N_c\int \frac {d{\bf p}}{(2\pi)^3}(n_i(T, \tilde{\mu})
-\bar{n}_i(T, \tilde{\mu})),
 \eeq
 where it is to be noted that the shifted chemical potential
 $\tilde {\mu }=\mu- 2g_{_V}\rho _i$ enters the distribution
functions instead of the naive one,  $\mu$.
 
Simply differentiating these equations with respect to $\mu$, one 
obtains $\chi_B(T,\mu)$.
It is noteworthy that  when $\mu\not=0$
 there arises a  coupling between $\chi_B$ and the scalar-density
  susceptibility $\chi_s$ owing to the non-vanishing ``vector-scalar
 susceptibility" $\chi_{_{VS}}$. They  are  defined by
\beq
\chi_s=-\frac {d \lla\bar q q\rra}{dm}
            =\beta \int d{\bf x}\lla\bar q(0,{\bf x})q
             (0,{\bf x})\bar {q }(0,{\bf 0})q(0,{\bf 0})\rra,
\eeq
\beq
\chi_{_{VS}}=\frac {\partial \lla\bar {q }q\rra}
{\partial \mu_B}
            =\beta \int d{\bf x}\lla
            \bar {q}(0,{\bf x})\gamma _0 q (0,{\bf x})
\bar {q }(0,{\bf 0})q(0,{\bf 0})\rra,
\eeq
respectively.  $\chi _s$ represents  the fluctuation of 
the order parameter of the chiral transition, and is  related with the 
 sigma meson propagator.
The differentiation leads to the coupled equation
\beq
\left(\begin{array}{cc}
1+2g_{_V}\chi _B^{(0)} & -\chi _{_{VS}}^{(0)}\\
4g_sg_{_V}\chi _{_{VS}}^{(0)} & 1-2g_s\chi_s^{(0)}\end{array}
\right)\left(\begin{array}{c}
\chi _B\\
-2g_s\chi _{_{VS}}\end{array}\right)=
\left(\begin{array}{c}\chi_B^{(0)}\\ 2g_s\chi _{_{VS}}^{(0)}\end{array}\right)
\eeq
where
$\chi _B^{(0)}(T,\mu_q)$
is the zero-th order baryon-number susceptibility given before,
\beq
\chi _{VS}^{(0)}=-2N_c\beta \sum_{i=1}^{N_f}\int {{d{\bf p}}\over {(2\pi)^3}}
\Bigl\{ n_i(1-n_i)-\bar {n}_i(1-\bar {n}_i)\Bigr\}
\eeq
the zero-th order vector-scalar one and
\beq
\chi _s^{(0)}&=& 2N_c\sum_{i=1}^{N_f} \int {{d{\bf p}}\over {(2\pi)^3}}
[\frac{p^2}{E_i^3}(1-n_i-\bar {n}_i) \nonumber \\ 
  \ & & + \beta \frac{M_i^2}{E_i^2}\{n_i(1-n_i)+
\bar {n}_i(1-\bar {n}_i)\}]
\eeq
 the scalar-density susceptibility in the zero-th order. 
 One should note that when $\mu =0$, then $\chi _{VS}^{(0)}$ vanishes.
  One can readily  obtain  $\chi _q$ and $\chi_s$ in terms of 
 $\chi ^{(0)}, \chi _{VS}^{(0)}$
  and $\chi_s^{(0)}$, so we do not write them down to save the space.


\subsection{(A) Finite density case $g_V=0$}

To see how the nature of the chiral transition can affect the behavior of
$\chi_B$, let us first take a simple case where $g_V$ is negligible.

The essential point lies in the fact that the distribution function
$n(T,\mu)=[\e^{\beta(E_p-\mu)}+1]^{-1}$  depends on $\mu$ not only explicitly
 but also  implicitly through the dynamical mass 
$M(T,\mu)$ in $E_p=\sqrt{M^2+p^2}$; hence
\beq
T\frac{\d n}{\d \mu}=n(1-n)-\frac{M}{E_p}\frac{\d M}{\d \mu}n(1-n).
\eeq
Thus
\beq
\chi_B(T, \mu)&=& 2N_c\sum_{i=1}^{N_f}
\beta\int {{d{\bf p}}\over {(2\pi)^3}}
\Bigl[\{ n_i(1-n_i)+\bar {n}_i(1-\bar {n}_i)\} \nonumber \\ 
 \ \ & &  - \frac{M_i}{E_{ip}}\frac{\d M_i}{\d \mu}\{n_i(1-n_i)-
\bar {n}_i(1-\bar {n}_i)\} \Bigr].
\eeq
The notable point is  the presence of the derivative
$\frac{\d M}{\d \mu}$. The constituent mass $M$ varies with the
quark condensate $\lla \bar{q}q\rra$ by Eq. (3.9), the order parameter of the chiral
transition.  We saw in \S 2 that the chiral transition at low temperatures is
likely to be of first order in the chemical potential direction.
It means that the derivative $\frac{\d M}{\d \mu}$ diverges at the critical
point at low temperatures, hence so does the susceptibility $\chi_B$. 

\subsection{(B) Zero-density case  with $g_V\not=0$ }

  Putting  $\mu=0$ into the expressions one gets\cite{qnumk} 
\beq
\chi _B={{\chi _B^{(0)}(T)}\over {1+2g_{_V}\chi _B^{(0)}(T)}},
\eeq
where
$\chi _B^{(0)}(T)$ is the susceptibility for the free-quark gas.
 The denominator of $\chi _B$ is essentially the inverse of the 
 propagator of the vector meson in the ring approximation
 at the vanishing four momenta. 
The above expression shows that
  $\chi _B$ is suppressed  by the vector coupling ( $g_{_V}>0$).
This is reasonable at least for a system with a finite $\mu $;
   because  the system becomes hard 
  to compress when the vector coupling is present, 
the number fluctuations will be suppressed\footnote{
 The interactions due to 
  vector fields  like $\omega $ or neutral $\rho $ mesons increases 
 the repulsion  between the constituents of the system.}
Recall also that  $\chi _B$ is proportional to 
  the compressibility $\kappa _T$. 

The comparison with the lattice data\cite{qnumk} shows that 
the vector coupling is rather small in the high temperature phase.
  It is interesting that this suppression of the vector coupling 
at high temperatures is consistent with the observation
   that the screening masses of the vector modes obtained in the 
 lattice simulations \cite{lattice} almost coincides with $2\pi T$,
 the lowest screening
  mass of the q-$\bar {\rm q}$ system in the chiral limit; a similar
result is also obtained in the instanton approach\cite{instanton}.

\section{Estimate of $g_{_V}$ and $g_{_S}$ in the lattice QCD} 

Boyd et al \cite{boyd} once extracted the effective coupling constants $g_{_V}$ and
 $g_{_S}$ from the lattice data using the expressions given in the NJL model
 as given above;
$\chi_B={\chi^{(0)}_B}/({1+g_{_V}\chi^{(0)}_B})$ and 
$\chi_S=\chi^{(0)}_S/({1-g_{_S}\chi^{(0)}_S})$.
They concluded that when $T\sim T_C$,
$g_{_S}$ is much bigger than $g_{_V}$; $g_{_S}\simeq 4g_{_V}$.
This result is consistent with 
our analysis and the behavior of the screening masses.

\section{Implications to phenomenology}

Kumagai, Miyamura and Sugitate\cite{kumagai} discussed the implications 
 of the baryon number susceptibility and the strangeness susceptibility
to the observables in the relativistic heavy-ion collisions.
They argued that in the stopping region where the chemical potential is
 large, large fluctuations of the baryon and the strangeness numbers 
may be a signature of the chiral transition.

Here we wish to 
also indicate that the large number fluctuations cause  those
 in the scalar channel (the sigma meson channel) at finite density.

The observability of the possible large density fluctuations caused as 
 a critical phenomenon 
 is discussed by several authors\cite{phenomena}.

\setcounter{equation}{0}  
\section{Summary and concluding remarks}

We have examined the baryon-number susceptibility $\chi_B$ as an
 observable which reflects the confinement-deconfinement and the chiral
 phase transitions in hot and/or dense hadronic matter.
   
The suppression of $\chi_B$ at low temperatures and steep rise around the
critical temperature as shown in the lattice QCD may be roughly 
attributed to the confinement-deconfinement transition.
Nevertheless, we have shown 
that such a behavior  of $\chi_B$ is also affected by the
chiral transition.

Since $\chi_B$ is a measure of the rate of the density fluctuation in the 
system,  the chiral transition at finite chemical potential 
especially leads to an interesting phenomenological consequence to 
$\chi_B$. When the vector coupling is small, the chiral transition 
at low temperatures is of first order in the density direction, which
 implies a divergent behavior of $\chi_B$, accordingly a huge density 
fluctuations. We have emphasized that such a large enhancement of the
fluctuation can be also expected  for the scalar density fluctuations
due to the scalar-vector mixing at finite density.
Such a large enhancement may leads to an enhancement of the
sigma-meson
 production
\footnote{As for the significance of the sigma meson for the chiral
transition at finite $T$ and/or $\rho_B$, see  \cite{physrep,hks}.}.
The above phenomena all have relevance to experiments to be done in RHIC
and LHC. 

We have indicated that the nature of the chiral transition as to the
first order or not etc is sensitively dependent on the strength of the
vector coupling. An analysis of the  lattice data suggests that the vector 
coupling is small in comparison with the scalar coupling at high temperature.

The susceptibility $\chi_B$ is nothing but 
the generalized susceptibility 
$\chi(\omega,k)$ at $\omega = k =0$.  One should examine $\chi(\omega,k)$
 in the whole region of $\omega $ and $k$ to get 
 more information about the vector correlations and the density fluctuations. 
 
In conclusion, we would like to thank the organizers of this workshop, 
especially, Prof. H. Suganuma, for inviting me to talk on 
the baryon-number susceptibility.

\end{document}